\documentstyle[12pt]{article}
\newenvironment{destaque}{\begin{quotation}\small\em}{\end{quotation}}

\begin{document}

\date{}
\title{A TEORIA DE RENORMALIZA\c{C}\~AO NO C\'ALCULO DOS POTENCIAIS ESCALAR
EL\'ETRICO E VETORIAL MAGN\'ETICO\\
(Renormalization theory in the electrostatic and vector potential
calculation)}
\author{{Wesley Spalenza and Jos\'e Alexandre Nogueira}\\
{\it Departamento de F\'{\i}sica, Centro de Ci\^encias Exatas,}\\
{\it Universidade Federal do Espirito Santo,}\\
{\it 29.060-900 - Vit\'oria-ES - Brasil,}\\
{\it E-mail: } spalenza@yahoo.com, nogueira@cce.ufes.br}
\maketitle

{\bf Resumo}

\begin{destaque}
Neste trabalho tentamos mostrar de uma maneira clara e simples as id\'{e}ias
fundamentais da Teoria de Renormaliza\c{c}\~{a}o. Neste intuito usamos dois
problemas bem conhecidos dos alunos de gradua\c{c}\~{a}o de Ci\^{e}ncias
Exatas, os c\'{a}lculos do potencial escalar el\'{e}trico e vetorial magn%
\'{e}tico de um fio infinito de carga e corrente el\'{e}trica,
respectivamente. Ainda, diferentes m\'{e}todos de regulariza\c{c}\~{a}o s%
\~{a}o usados (cut-off, dimensional e fun\c{c}\~{a}o zeta) e o aparecimento
do par\^{a}metro de escala \'{e} discutido.\newline
Palavra chave: renormaliza\c{c}\~{a}o, regulariza\c{c}\~{a}o e par\^{a}metro
de escala.
\end{destaque}

\vspace{1cm}

{\bf Abstract}

\begin{destaque}
In this work we attempt to show in a clear and simple manner the fundamental
ideas of the Renormalization Theory. With that intention we use two
well-known problems of the Physic and Engeneering undergraduate students,
the calculation of the electrostatic and vector potential of a infinite line
charge density and current, respectively. We still employ different
regularization methods (cut-off, dimensional and zeta function) and the
arising of the scale parameter is consider.\newline
\end{destaque}

\section{Introdu\c{c}\~ao}

\hspace{.5cm} Nos dias atuais a Teoria Qu\^antica de Campos \'e largamente
empregada em diversas \'areas da f\'{\i}sica, tais como, altas energias,
mec\^anica estat\'{\i}stica, mat\'eria condensada, etc. Sendo a Teoria
Qu\^antica de Campos fundamentalmente de aspectos perturbativos, ela sofre
de pesados problemas de diverg\^encias. O tratamento destas diverg\^encias
tem sido um enorme desafio para os f\'{\i}sicos. A natureza matem\'atica do
problema \'e bem conhecida. Diverg\^encias ocorrem nos c\'alculos
perturbativos porque duas distribui\c{c}\~oes n\~ao podem ser multiplicadas
em um mesmo ponto. V\'arios m\'etodos tem sido propostos para solucionar
este problema. Entretanto somente \'e poss\'{\i}vel eliminar estes infinitos
de uma maneira f\'{\i}sica e consistente por absorv\^e-los nos par\^ametros
livres da teoria (massa e constante de acoplamento).

O procedimento usual para sanar o problema das diverg\^encias \'e empregar
um m\'etodo de regulariza\c{c}\~ao (cut-off, dimensional, zeta, etc ),
tornando a teoria finita atrav\'es do uso de um regulador (par\^ametro de
regulariza\c{c}\~ao) a fim de isolar as diverg\^encias e, ent\~ao,
restabelecer a teoria original com a elimina\c{c}\~ao do regulador usando
uma prescri\c{c}\~ao de renormaliza\c{c}\~ao, subtra\c{c}\~ao dos p\'olos ou
adi\c{c}\~ao de contra-termos.

De maneira geral o entendimento do procedimento de renormaliza\c{c}\~ao
empregado fica prejudicado devido a complexidade da Teoria Qu\^antica de
Campos. A fim de contornar esta dificuldade, vamos tratar aqui de dois
problemas simples e bem conhecidos por qualquer aluno de gradua\c{c}\~ao em
f\'{\i}sica e possivelmente dos demais cursos da \'area de Ci\^encias Exatas.

Os problemas aos quais nos referimos \'e o da determina\c{c}\~ao do
potencial escalar el\'etrico e do potencial vetor magn\'etico de um fio
infinito de carga e de corrente, respectivamente. Tais problemas, de um modo
geral parecem amb\'{\i}guos para os alunos, pois escondido neles existe um
procedimento de renormaliza\c{c}\~ao, como apontou Hans em seu artigo [1].
Uma maneira encontrada para se evitar diretamente as diverg\^encias nos
c\'alculos dos potenciais, \'e primeiramente determinar os campos
el\'etricos e magn\'etico e em seguida calcular os potenciais escalar
el\'etrico e vetorial magn\'etico do fio infinito.

O artigo est\'a organizado com segue. Na se\c{c}\~ao-2 tratamos do c\'alculo
do potencial escalar el\'etrico de um fio infinito com densidade linear de
carga $\lambda$ e do potencial vetor magn\'etico de um fio infinito de
corrente constante, que nos conduzir\'a a uma integral divergente. Nas se\c{c%
}\~oes 3, 4 e 5 n\'os regularizamos a integral divergente obtida na se\c{c}%
\~ao anterior usando os m\'etodos, cut-off [3], dimensional [4] e fun\c{c}%
\~ao zeta [5] respectivamente. Na se\c{c}\~ao-6 usando as prescri\c{c}\~oes
de renormaliza\c{c}\~ao, determinamos os potenciais renormalizados,
discutimos o par\^ametro de escala e apresentamos as id\'eias b\'asicas da
Teoria de Renormaliza\c{c}\~ao em Teoria Qu\^antica de Campos.

\section{Potencial Escalar El\'etrico e Potencial Vetor Magn\'etico}

\hspace{.5cm} O potencial escalar el\'etrico $\Phi(\vec{r})$ gerado por um
fio infinito com densidade linear de carga $\lambda$ em um ponto qualquer do
espa\c{c}o exceto no fio \'e dado por [2-3]
\begin{eqnarray}
\Phi(\vec{r})=\frac{\lambda}{4\pi\varepsilon_{0}}\int_{-\infty}^{\infty}%
\frac{dz} {\sqrt{z^{2}+\rho^{2}}},
\end{eqnarray}
onde temos colocado o fio sobre o eixo z e $\rho$ \'e a dist\^ancia do ponto
ao fio, coordenada radial cil\'{\i}ndrica.

O potencial vetor magn\'etico $\vec{A}(\vec{r})$ produzido por um fio
infinito de corrente el\'etrica constante $i$, \'e dado por [3]
\begin{eqnarray}
\vec{A}(\vec{r})=\frac{\mu_{0}i}{4\pi}\int_{-\infty}^{\infty}\frac{dz}{\sqrt{%
z^{2}+\rho^{2}}}\hat{k},
\end{eqnarray}
onde temos usando a mesma geometria anterior.

Uma an\'alise dimensional da integral
\begin{eqnarray}
I=\int_{-\infty}^{\infty}\frac{dz}{\sqrt{z^{2}+\rho^{2}}},
\end{eqnarray}
que aparece nas equa\c{c}\~oes dos potenciais, mostra que ela \'e
adimensional e portanto sofre de uma diverg\^encia logar\'{\i}tmica.

Assim, vemos que para estes dois problemas simples devemos empregar um
procedimento de renormaliza\c{c}\~ao a fim de obtermos os potenciais
renormalizados, isto \'e, "observados" (a difer\^en\c{c}a de potencial entre
dois pontos, pois ele \'e uma grandeza relativa e n\~ao absoluta).

A fim de tornar a teoria finita e assim manuze\'avel, devemos empregar um
m\'etodo de regulariza\c{c}\~ao. Isto vai nos permitir separarmos a parte
finita da divergente. Por\'em, a teoria fica dependente de um par\^ametro de
regulariza\c{c}\~ao e uma prescri\c{c}\~ao de renormaliza\c{c}\~ao dever\'a
se empregada para restabelecermos a teoria original. Vamos utilizar
diferentes m\'etodos de regulariza\c{c}\~ao e mostrar que, embora cada um
forne\c{c}a um resultado diferente, a teoria final, isto \'e, renormalizada
(f\'{\i}sica) \'e independente do m\'etodo de regulariza\c{c}\~ao usado.

\section{Cut-off}

\hspace{.5cm} Esse m\'etodo de regulariza\c{c}\~ao se baseia no emprego de
um corte nos limites da integral, isto \'e, trocamos o limite infinito por
um valor finito $\Lambda$ (par\^ametro regularizador).

Com a inclus\~ao do corte tornamos a teoria finita, por\'em dependente de $%
\Lambda$. Portanto, para restabelecermos a teoria original, devemos ao final
tomar o limite com $\Lambda$ tendendo a infinito.

Na integral da eq.(3) vamos introduzir um corte

\begin{eqnarray}
I_{\Lambda}=\int_{0}^{\Lambda}\frac{dz}{\sqrt{z^{2}+\rho^{2}}}.
\end{eqnarray}

Uma vez que tomaremos o limite, \'e conveniente obtermos o resultado da
integral da eq.(4) em pot\^encias de $\Lambda$ e de $\frac{1}{\Lambda}$ de
forma a permitir a separa\c{c}\~ao do(s) p\'olo(s) da parte finita. Vamos
dividir a integral da eq.(4) em duas partes

\begin{eqnarray}
I_{\Lambda}=\int_{0}^{\rho}\frac{dz}{\rho\sqrt{\frac{z^{2}}{\rho^{2}}+1}} +
\int_{\rho}^{\Lambda}\frac{dz}{z\sqrt{\frac{\rho^{2}}{z^{2}}+1}},
\end{eqnarray}
para considerarmos os casos em que $z<\rho$ e $z>\rho$. Realizando as
expans\~oes em s\'erie de Taylor dos integrandos da eq. (5) e depois
integrando termo a termo obtemos

\[
I_{\Lambda }=C+\ln \left( \frac{\Lambda }{\rho }\right) +O\left( \frac{1}{%
\Lambda ^{2}}\right) ,
\]
onde $C$ \'{e} uma constante.

Podemos observar que quando tentamos restabelecer a teoria original, ou
seja, tomamos o limite de $\Lambda$ tendento a infinito, presenciamos uma
diverg\^encia logar\'{\i}tmica, como j\'a esperavamos.

\section{Regulariza\c{c}\~ao Dimensional}

\hspace{.5cm} Este m\'etodo de regulariza\c{c}\~ao consiste em modificar a
dimens\~ao da integral atrav\'es de uma continua\c{c}\~ao anal\'{\i}tica de
forma a torn\'a-la finita. Consegue-se isto trocando a dimens\~ao do
diferenciando por uma outra complexa, atrav\'es da inclus\~ao de um
par\^ametro regularizador complexo, $\omega$

\begin{eqnarray}
I(\rho,\omega)=\int_{-\infty}^{\infty}\frac{d^{1-\omega}z}{\sqrt{z^{2}+
\rho^{2}}}.
\end{eqnarray}

A integral (7) agora \'e finita e pode ser realizada usando a rela\c{c}\~ao
[4]

\[
\int_{-\infty }^{\infty }\left( k^{2}+a^{2}\right) ^{-\alpha }d^{m}k=\pi ^{%
\frac{m}{2}}\frac{\Gamma (\alpha -\frac{m}{2})}{\Gamma (\alpha )}\left(
a^{2}\right) ^{\frac{m}{2}-\alpha },
\]
obtendo
\[
I(\rho ,\omega )=\pi ^{\frac{-\omega }{2}}\Gamma \left( \frac{\omega }{2}%
\right) (\rho )^{-\omega }.
\]

Para separarmos a parte finita da diverg\^{e}nte quando $\omega $ vai a
zero, vamos fazer uma expans\~{a}o em pot\^{e}ncias de $\omega $ da eq.(9),
para isto usamos para $|\omega |\ll 1$ as seguintes rela\c{c}\~{o}es
\[
\Gamma \left( \frac{\omega }{2}\right) =\frac{2}{\omega }-\gamma +O(\omega )
\]
e
\[
\rho ^{-\omega }=1-\frac{\omega }{2}\ln (\rho ^{2})+O(\omega ^{2}),
\]
onde $\gamma $ \'{e} o n\'{u}mero de Euler. Ent\~{a}o temos
\[
I(\rho ,\omega )=\pi ^{-\frac{\omega }{2}}\left[ \frac{2}{\omega }-\gamma
-\ln \left( \frac{\rho ^{2}}{\mu ^{2}}\right) +O(\omega )\right] ,
\]
onde temos inclu\'{i}do um par\^{a}metro de escala $\mu $ com dimens\~{a}o
de comprimento, a fim de tornar o logaritmando adimensional.

\section{Regulariza\c{c}\~ao por Fun\c{c}\~ao Zeta}

\hspace{.5cm} A fun\c{c}\~ao zeta generalizada associada a um operador $M$,
\'e definida como
\begin{eqnarray}
\zeta_{M}(s)=\sum_{i}\lambda_{i}^{-s},
\end{eqnarray}
onde $\lambda_i$, s\~ao os auto-valores do operador $M$ e $s$ um par\^ametro
complexo

Definimos, para o nosso caso, a fun\c{c}\~ao zeta como

\[
\zeta (s+1/2)=\int_{-\infty }^{\infty }\left( \frac{z^{2}}{\mu ^{2}}+\frac{%
\rho ^{2}}{\mu ^{2}}\right) ^{-s-1/2}d\left( \frac{z}{\mu }\right)
\]
e a integral (3) fica
\[
I(\rho ,s)=\zeta (s+1/2).
\]
O par\^{a}metro de escala $\mu $, com dimens\~{a}o de comprimento foi
inclu\'{i}do para tornar a fun\c{c}\~{a}o zeta admensional para todo $s$.%
\newline
Usando a rela\c{c}\~{a}o (8) obtemos
\[
\zeta (s+1/2)=\sqrt{\pi }\frac{\Gamma (s)}{\Gamma (s+1/2)}\left( \frac{\rho
^{2}}{\mu ^{2}}\right) ^{-s}
\]
que com a aproxima\c{c}\~{a}o
\[
2\sqrt{\pi }\frac{\Gamma (s)}{\Gamma (s-1/2)}\approx -\frac{1}{s},
\]
para $|s|\ll 1$, temos
\[
\zeta (s+1/2)=-\frac{\left( \frac{\rho ^{2}}{\mu ^{2}}\right) ^{-s}}{%
2s(s-1/2)}.
\]

A continua\c{c}\~{a}o anal\'{i}tica para s igual a zero da eq.(18) \'{e}
obtida multiplicando a equa\c{c}\~{a}o por s e em seguida derivando em $s=0$
[5]. Assim
\[
\Phi (\vec{r})=\frac{\lambda }{2\pi \varepsilon _{0}}-\frac{\lambda }{2\pi
\varepsilon _{0}}\ln \left( \frac{\rho }{\mu }\right) ,
\]
\[
\vec{A}(\vec{r})=\frac{\mu _{0}i}{2\pi }\hat{k}-\frac{\mu _{0}i}{2\pi }\ln
\left( \frac{\rho }{\mu }\right) \hat{k}.
\]

\section{Condi\c{c}\~oes de Renormaliza\c{c}\~ao}

\hspace{.5cm} Como podemos observar os potenciais obtidos atrav\'es dos
resultados dados pelas eq.(6) e (12) s\~ao ainda divergentes, portanto,
devemos lan\c{c}ar m\~ao de uma prescri\c{c}\~ao de renormaliza\c{c}\~ao a
fim de eliminar a parte divergente (p\'olo).

Como prescri\c{c}\~{a}o de renormaliza\c{c}\~{a}o usaremos a condi\c{c}%
\~{a}o f\'{i}sica, de que os potenciais n\~{a}o s\~{a}o grandezas absolutas
e sim relativas, isto \'{e}, somente diferen\c{c}as de potenciais podem ser
observadas. Assim, usando as eq.(6) e (12) obtemos
\[
\Phi (\vec{r})-\Phi (\vec{r_{0}})=\frac{\lambda }{2\pi \varepsilon _{0}}\ln
\left( \frac{\rho _{0}}{\rho }\right)
\]
e
\[
\vec{A}(\vec{r})-\vec{A}(\vec{r_{0}})=\frac{\mu _{0}i}{2\pi }\ln \left(
\frac{\rho _{0}}{\rho }\right) \hat{k}
\]
Agora tomando o potencial nulo no ponto de refer\^{e}ncia $\vec{r_{0}}$,
temos
\[
\Phi _{R}(\vec{r})=\frac{\lambda }{2\pi \varepsilon _{0}}\ln \left( \frac{%
\rho _{0}}{\rho }\right)
\]
e
\[
\vec{A}_{R}(\vec{r})=\frac{\mu _{0}i}{2\pi }\ln \left( \frac{\rho _{0}}{\rho
}\right) \hat{k}.
\]

Note que o ponto de refer\^encia $\vec{r_{0}}$ \'e completamente
arbitr\'ario.

Embora os resultados obtidos nas eq.(19) e (20) sejam finitos, eles ainda
n\~ao representam os resultados f\'{\i}sicos, pois n\~ao sabemos se o que
retiramos da parte divergente foi mais que o necess\'ario. Uma renormaliza\c{%
c}\~ao finita deve ser realizada para que os potenciais obtidos sejam
aqueles que representem a f\'{\i}sica do problema.

Novamente usando a diferen\c{c}a de potencial como condi\c{c}\~ao de
renormaliza\c{c}\~ao, obtemos das eq.(19) e (20) os mesmos resultados
obtidos nas eq.(23) e (24)

\'E importante comentarmos a presen\c{c}a do par\^ametro de escala $\mu$ nas
eq.(12), (19) e (20).

A prescri\c{c}\~{a}o de renormaliza\c{c}\~{a}o usada aqui fornece
imediatamente o resultado f\'{i}sico, isto \'{e}, o potencial no ponto $\vec{%
r}$ medido em rela\c{c}\~{a}o aquele medido no ponto de refer\^{e}ncia $\vec{%
r_{0}}$. Se desejassemos como primeira etapa obter um resultado finito para
as eq.(6) e (12) poder\'{i}amos usar como prescri\c{c}\~{a}o a subtra\c{c}%
\~{a}o do termo divergente (p\'{o}lo). Na eq.(6) a fim de separarmos a parte
divergente da finita devemos multiplicar e dividir o logaritimando por um
par\^{a}metro arbitr\'{a}rio finito, o par\^{a}metro de escala $\mu $.
\[
I(\rho ,\mu ,\Lambda )=C-\left[ \ln \left( \frac{\rho }{\mu }\right) -\ln
\left( \frac{\Lambda }{\mu }\right) \right] +O\left( \frac{1}{\Lambda ^{2}}%
\right) .
\]
Agora usando como prescri\c{c}\~{a}o a subtra\c{c}\~{a}o do p\'{o}lo,
obtemos, para o cut-off
\[
\Phi (\vec{r})=\frac{\lambda }{2\pi \varepsilon _{0}}\ln \left( \frac{\rho }{%
\mu }\right) +\frac{\lambda }{2\pi \varepsilon _{0}}C,
\]
e para a dimensional
\[
\Phi (\vec{r})=\frac{\lambda }{2\pi \varepsilon _{0}}\ln \left( \frac{\rho }{%
\mu }\right) +\frac{\gamma }{2\pi \varepsilon _{0}}.
\]

Ent\~ao, notamos que no caso da regulariza\c{c}\~ao dimensional e zeta, esta
separa\c{c}\~ao j\'a foi realizada de alguma forma escondida dentro dos
procedimento usados.

Uma maneira mais elegante e formal de introduzimos o par\^{a}metro de
escalar \'{e} fazendo com que a integral inicial (3) seja adimensional, isto
\'{e},
\[
I=\int_{-\infty }^{\infty }\frac{d\left( \frac{z}{\mu }\right) }{\sqrt{\frac{%
z^{2}}{\mu ^{2}}+\frac{\rho ^{2}}{\mu ^{2}}}}.
\]
E desta forma tornando a eq.(7) adimensional para qualquer $\omega $.

\'{E} claro que a continua\c{c}\~{a}o anal\'{i}tica usada no m\'{e}todo da
fun\c{c}\~{a}o zeta \'{e} a prescri\c{c}\~{a}o de renormaliza\c{c}\~{a}o
necess\'{a}ria para se obter o resultado finito e \'{e} equivalente a subtra%
\c{c}\~{a}o do p\'{o}lo. Isso fica claro se tivessemos realizado a
expans\~{a}o em s\'{e}rie de Laurent da eq.(18)
\[
I(\rho ,s)=\frac{a_{-1}}{s}+\ln \left( \frac{\rho }{\mu }\right) +O(s),
\]
onde $a_{-1}$ \'{e} o res\'{i}duo.

Note que os resultados das eq.(19),(26) e (27) diferem por uma constante e
s\~ao dependentes do par\^ametro de escala. Como j\'a dissemos, embora os
resultados destas equa\c{c}\~oes sejam finitos eles ainda n\~ao representam
a f\'{\i}sica da teoria. Isto \'e obvio, pois, n\~ao podemos ter os
resultados f\'{\i}sicos (observados) dependentes do m\'etodo de regulariza\c{%
c}\~ao. Uma renormaliza\c{c}\~ao finita deve ser feita para ajustar os
potenciais obtidos aqueles observados (diferen\c{c}as). Esta condi\c{c}\~ao
de renormaliza\c{c}\~ao nos permite escrever os potenciais em fun\c{c}\~ao
daqueles observados em um determinado ponto. Ela tamb\'em permite que o
par\^ametro de escala seja escrito em fun\c{c}\~ao do ponto de refer\^encia $%
\rho_{0}$. \'E claro que o ponto de refer\^encia \'e arbitr\'ario e portanto
tamb\'em o par\^ametro de escala.

Agora estamos aptos a sintetizar como funciona a renormaliza\c{c}\~ao. Os
potenciais dados pelas eq.(6), (12) e (19), n\~ao s\~ao aqueles f\'{\i}sicos
(observ\'aveis) sendo at\'e mesmo divergentes. Para torn\'a-los aqueles
observados devemos ajust\'a-los. Assim, medimos (na verdade aqui definimos
um valor qualquer, em geral zero) o potencial em um ponto de refer\^encial
qualquer $\vec{r_{0}}$ que no caso da Teoria Qu\^antica de Campos \'e
chamado ponto de renormaliza\c{c}\~ao ou subtra\c{c}\~ao. Por fim escrevemos
o potencial f\'{\i}sico (observado) como fun\c{c}\~ao daquele medido no
ponto de refer\^encia (ponto de renormaliza\c{c}\~ao). Este procedimento
ent\~ao absorve a diverg\^encia do potencial original n\~ao f\'{\i}sico.

Em resumo:

i) Potencial original n\~ao f\'{\i}sico
\begin{eqnarray}
\Phi_d(\vec{r})=D + C + \Phi_F(\vec{r}),
\end{eqnarray}
onde D \'e o termo divergente separado por um m\'etodo qualquer de regulariza%
\c{c}\~ao, e C \'e uma constante que depende do m\'etodo de regulariza\c{c}%
\~ao e $\Phi_F(\vec{r})$ \'e o potencial.

ii) Potencial medido no ponto de refer\^encia (renormaliza\c{c}\~ao)
\begin{eqnarray}
\Phi_{0}=D + C + \Phi_F(\vec{r_{0}}).
\end{eqnarray}
Neste caso para $\Phi_{0}$ \'e determinado um valor arbitr\'ario e n\~ao
realmente medido

Agora escrevemos
\begin{eqnarray}
D + C = \Phi_{0}-\Phi_F(\vec{r_{0}})
\end{eqnarray}
e substituindo na eq.(31), fica
\begin{eqnarray}
\Phi_R(\vec{r})=\Phi(\vec{r})-\Phi(\vec{r_{0}})+\Phi_{0},
\end{eqnarray}
onde $\Phi_R(\vec{r})$ \'e o potencial renormalizado.

Note que mesmo no caso de um m\'etodo de regulariza\c{c}\~ao que forne\c{c}a
um resultado finito, ainda temos de ajustar este resultado aquele
f\'{\i}sico.

Finalmente, podemos analizar como funciona a renormaliza\c{c}\~ao na Teoria
Qu\^antica de Campos. A teoria original depende de alguns par\^ametros em
geral divergentes, tais como $m$ e $\lambda$. Tais par\^ametro n\~ao
representam a massa $(m)$ e a constante de acoplamento $\lambda$ observados
da teoria e sim s\~ao ajustando atrav\'es das condi\c{c}\~oes de renormaliza%
\c{c}\~ao a estas quantidades f\'{\i}sicas renormalizadas, medidas em caso
de teorias realistas, ou definidas no caso de teorias n\~ao realistas, em um
determinado ponto, chamado ponto de renormaliza\c{c}\~ao ou subtra\c{c}\~ao.
Este ponto, pode ser o quadri-momento da teoria ou um determinado estado do
sistema, em geral o de menor energia, ou estado de v\'acuo, embora qualquer
ponto seja t\~ao bom quanto outro, isto \'e, o ponto de renormaliza\c{c}\~ao
\'e arbitr\'ario.

Escrevendo agora a teoria original em fun\c{c}\~ao n\~ao mais dos
par\^ametros originais $m$ e $\lambda$ e sim das quantidades f\'{\i}sicas
renormalizadas ("observadas") $m_{R}$ e $\lambda_{R}$, as diverg\^encias
s\~ao absorvidas de forma semelhante ao que ocorreu com o potencial.

Uma maneira alternativa usada \'e tomar os par\^ametros $m$ e $\lambda$ da
teoria original como sendo realmente aquele observados (renormalizados) e
absorver as diverg\^encias da teoria em contra-termos $\delta$$m$ e $%
\delta\lambda$ inclu\'{\i}dos na teoria. Tais contra-termos, \'e claro,
devem ser de termos de mesma pot\^encia nos campos que aqueles de $m$ e $%
\lambda$. Ent\~ao, usando as condi\c{c}\~oes de renormaliza\c{c}\~ao os
contra-termos s\~ao determinados de forma a anular as diverg\^encias e
fornecer a f\'{\i}sica da teoria.

\section{Conclus\~ao}

\hspace{.5cm} Atrav\'es de um exemplo simples do c\'alculo dos potenciais
escalar e vetorial de um fio infinito de carga e de corrente,
respectivamente, podemos apresentar as diverg\^encias que sofrem algumas
teorias, os m\'etodos usados para lidar com estas diverg\^encias
(separ\'a-los da parte finita) e o procedimento usado para tornar tais
teorias em teorias f\'{\i}sicas (renormaliza\c{c}\~ao).

\end{document}